\magnification\magstephalf
\overfullrule 0pt
\input epsf

\font\rfont=cmr9 at 9 true pt
\def\ref#1{$^{\hbox{\rfont {[#1]}}}$}


\font\tenbfit=cmbxti10
\font\sevenbfit=cmbxti10 at 7pt
\font\fivebfit=cmbxti10 at 5pt
\newfam\bfitfam 
\textfont\bfitfam=\tenbfit  \scriptfont\bfitfam=\sevenbfit
\scriptscriptfont\bfitfam=\fivebfit

\font\twelvebf=cmbx12

\font\tenrm=cmr10 scaled\magstep0

\font\tenbfit=cmbxti10
\font\sevenbfit=cmbxti10 at 7pt
\font\fivebfit=cmbxti10 at 5pt
\newfam\bfitfam 
\textfont\bfitfam=\tenbfit  \scriptfont\bfitfam=\sevenbfit
\scriptscriptfont\bfitfam=\fivebfit

\font\tenbit=cmmib10
\newfam\bitfam
\textfont\bitfam=\tenbit%

\font\tenmbf=cmbx10
\font\sevenmbf=cmbx7
\font\fivembf=cmbx5
\newfam\mbffam
\textfont\mbffam=\tenmbf \scriptfont\mbffam=\sevenmbf
\scriptscriptfont\mbffam=\fivembf

\font\tenbsy=cmbsy10
\newfam\bsyfam 
\textfont\bsyfam=\tenbsy%

   
\def\e{\epsilon}

\def\pd {\partial}
\def\pmb#1{\setbox0=\hbox{#1}
 \kern.05em\copy0\kern-\wd0 \kern-.025em\raise.0433em\box0 }

\def \half {{\scriptstyle {1 \over 2}}}

 %


\def\boxit#1{\vbox{\hrule\hbox{\vrule\kern1pt\vbox
{\kern1pt#1\kern1pt}\kern1pt\vrule}\hrule}}

\def\h{\hfill\break}
\parskip=6pt
\parindent=0pt
\hsize=17truecm\hoffset=-5truemm
\voffset=-1truecm\vsize=24.5truecm
\def\footnoterule{\kern-3pt
\hrule width 17truecm \kern 2.6pt}


\catcode`\@=11 

\def\nolabels{\def\wrlabeL##1{}\def\eqlabeL##1{}\def\reflabeL##1{}}
\def\writelabels{\def\wrlabeL##1{\leavevmode\vadjust{\rlap{\smash%
{\line{{\escapechar=` \hfill\rlap{\sevenrm\hskip.03in\string##1}}}}}}}%
\def\eqlabeL##1{{\escapechar-1\rlap{\sevenrm\hskip.05in\string##1}}}%
\def\reflabeL##1{\noexpand\llap{\noexpand\sevenrm\string\string\string##1}}}
\nolabels
\global\newcount\refno \global\refno=1
\newwrite\rfile
\def\defref{$^{{\hbox{\rfont [\the\refno]}}}$\nref}
\def\nref#1{\xdef#1{\the\refno}\writedef{#1\leftbracket#1}%
\ifnum\refno=1\immediate\openout\rfile=refs.tmp\fi
\global\advance\refno by1\chardef\wfile=\rfile\immediate
\write\rfile{\noexpand\item{#1\ }\reflabeL{#1\hskip.31in}\pctsign}\findarg}
\def\findarg#1#{\begingroup\obeylines\newlinechar=`\^^M\pass@rg}
{\obeylines\gdef\pass@rg#1{\writ@line\relax #1^^M\hbox{}^^M}%
\gdef\writ@line#1^^M{\expandafter\toks0\expandafter{\striprel@x #1}%
\edef\next{\the\toks0}\ifx\next\em@rk\let\next=\endgroup\else\ifx\next\empty%
\else\immediate\write\wfile{\the\toks0}\fi\let\next=\writ@line\fi\next\relax}}
\def\striprel@x#1{} \def\em@rk{\hbox{}} 
\def\lref{\begingroup\obeylines\lr@f}
\def\lr@f#1#2{\gdef#1{\defref#1{#2}}\endgroup\unskip}
\newwrite\lfile
{\escapechar-1\xdef\pctsign{\string\%}\xdef\leftbracket{\string\{}
\xdef\rightbracket{\string\}}}

\def\writestop{\def\writestoppt{\immediate\write\lfile{\string\p
ageno%
\the\pageno\string\startrefs\leftbracket\the\refno\rightbracket%
\string\def\string\secsym\leftbracket\secsym\rightbracket%
\string\secno\the\secno\string\meqno\the\meqno}\immediate\closeout\lfile}}
\def\writestoppt{}\def\writedef#1{}
\catcode`\@=12 

\input amssym.def
\input amssym.tex
\centerline{\twelvebf INTRODUCTION TO}
\vskip 5truemm
\centerline{\twelvebf EQUILIBRIUM THERMAL FIELD THEORY
\footnote{${}^{*}$}{{\tenrm 
Lectures given in
August 1998 at the University of Regensburg, based on a course given in
February 1997 at the IX Jorge Andr\'e Swieca Summer School, Campos do
Jord\~ao, Brazil}}}
\bigskip
\centerline{P V Landshoff}
\centerline{DAMTP, University of Cambridge}
\centerline{pvl@damtp.cam.ac.uk}
\vskip 30pt
\leftline{{\bf Abstract}}
\medbreak
Within the next few years experiments at RHIC and the LHC will seek
to create in the laboratory a quark-gluon plasma, the phase of matter through
which the Universe passed very early in its life. It is believed that the
plasma will survive long enough to reach thermal equilibrium.
I give an introduction to the formalism of thermal field theory, the
combination of statistical mechanics and quantum field theory needed to
describe the plasma in thermal equilibrium,  in a
way that tries to keep close to the physics it describes.

\vskip 30pt
{\bf Introduction}
\medbreak
Thermal field theory is a combination of quantum field theory and
statistical mechanics.  This means that it is both difficult and
interesting.  The reason that we study it is that we want to describe
the quark-gluon plasma, the phase that matter is believed to take
above some critical temperature $T_c$.  Lattice calculations
suggest\defref\lattice{
For a recent review, see A Ukawa,
Nucl Phys Proc Suppl 53 (1997) 106
} that $T_c$ is about $100\; {\rm MeV}$, or $10^{12}$K.
In the plasma phase the quarks and gluons are deconfined; they can
move rather freely through the whole plasma.  This is the phase to
which the universe evolved soon after the big bang, and before the end of
the century experiments at the new collider RHIC will try to re-create
it in the laboratory, by making gold nuclei collide together head-on
and dump their kinetic energy into a small volume. Similar experiments,
at much higher energy, are planned later  for the LHC at CERN.

There is an obvious question: if a plasma is indeed produced, how will
we know it? As yet there is no simple answer. There are estimates\defref\eq{
K J Eskola nd X N Wang, Physical Review D49 (1994) 1284
},
necessarily based on very crude non-equilibrium theory, that suggest
that the plasma will survive for a time long enough that it reaches
thermal equilibrium before it eventually decays back into ordinary matter.
So far, it is only equilibrium thermal field theory that is well formulated,
and my lectures concentrate on this.  For more
information, the book by Le~Bellac\defref\lebellac{
M Le Bellac, {\sl Thermal field theory}, Cambridge University Press (1996)
} nowadays
is the standard text, though the older book by Kapusta\defref\kapusta{
J Kapusta, {\sl Finite-temperature field theory}, Cambridge University Press
(1989)
}
is still valuable, as is the classic Physics Report by Landsman and
van Weert\defref\lvw{
N P Landsman and C G van Weert, Physics reports 145 (1987) 142
}.  As I want my description to stay as
close as possible to physics I will develop the theory using operators
rather than path integrals, and mostly I will use the so-called real-time
formalism. In thermal field theory it is very important to pay attention to
the physics: some of the clever mathematical formalism that has been 
developed is very powerful for calculating certain quantitities, but it
is all too easy to misuse it if one does not keep a close eye on the physics
it is supposed to describe.

Because in relativistic theory particles are continually being created
and destroyed, it is appropriate to use the grand ensemble formalism
of statistical mechanics. The grand ensemble consists of a very large
number of copies of the system under study. The energy and other conserved
quantum numbers, such as total lepton and baryon number, vary from
system to system in the ensemble, but their totals over the whole ensemble
are fixed, and therefore their average values per system are fixed.
If the system is large enough,
fluctuations about the average values are very small, and so the grand
ensemble formalism does effectively 
achieve the desired physical situation in which
the conserved quantum numbers are fixed for each system. In
practice we do not specify the ensemble in terms of their values;
rather we use related quantities, the temperature and various chemical
potentials. These arise as Lagrange multipliers when we maximise the
appropriate quantity in order to calculate the most probable configuration
of the ensemble. This mathematics leads to the grand partition function
$$
Z = \sum_{i} \langle i | e^{-\beta (H - \mu N)} |i \rangle \eqno{(1)}
$$
from which the macroscopic properties of the system in thermal
equilibrium are calculated.
Here $\beta$ is the inverse temperature, $\beta = 1/k_B T$, and
usually we use units in which Boltzmann's constant $k_B = 1$.  The
system's Hamiltonian is $H$ and $N$ is a conserved quantum number,
such as baryon number, with $\mu$ the corresponding chemical
potential.  In the case of several conserved quantum mumbers, 
$\mu N$ is replaced with $\textstyle{\sum_{\alpha}}\, \mu_{\alpha} N_{\alpha}$.

The states $|i\rangle$ are a complete orthonormal set of physical states of
the system.  In scalar field theory all states are physical and so
$$
Z = {\rm tr} \; e^{-\beta (H- \mu N)}\eqno{(2)}
$$
which is invariant under changes in the choice of orthonormal basis of
states.  In the case of gauge theories there are unphysical states,
for example longitudinally-polarised photons or gluons, which must be
excluded from the summation in (1).  So then
$$
Z = {\rm tr} \;{\Bbb P}\; e^{-\beta (H-\mu N)} \eqno{(3)}
$$
where ${\Bbb P}$ is a projection operator that removes unphysical states.  The
presence of ${\Bbb P}$ can make things more complicated, and so to begin
with I will consider scalar field theory, where it is not needed.

All the macroscopic properties of the system in thermal equilibrium
may be calculated from $Z$.  In particular, for a system that is so
large that its surface energy is negligible compared with its volume
energy, the equation of state is
$$
PV = T \; \log\; Z \eqno{(4)}
$$
Because we do not have sufficient knowledge of each system of the ensemble
to specify which system state $|i\rangle$ it is in, we use a density
matrix $\rho$ to describe the system: in thermal equilibrium
$$
\rho =Z^{-1} e^{-\beta (H-\mu N)}
\eqno(5a)
$$
The ``thermal average'' of an observable corresponding to an
operator $Q$, that is its average over all the systems of the ensemble, is
$$
<Q> =  {\rm tr}\, Q\rho  = Z^{-1}\; {\rm tr}\; Q e^{-\beta (H-\mu N)}  
\eqno{(5b)}
$$

Notice that, throughout, all operators and all states
are familiar zero-temperature
ones.  The temperature enters only in the exponential, which
defines how the zero-temperature states are combined together to define
the statistical-mechnical ensemble used to calculate
the thermal averages of the zero-temperature operators.
\bigskip
{\bf Noninteracting scalar bosons}

For some systems of bosons there is no conserved quantum number $N$
and therefore no corresponding Lagrange multiplier, that is there is no
chemical potential $\mu$.
For example, in the case of a heat bath of photons there is no
constraint on their total number.  Then in the scalar-field-theory
case $Z$ is just ${\rm tr}\; e^{-\beta H}$. 

Consider
first the case where interactions among the particles are so small
that, once the system has thermalised, they can be neglected.
This is a reasonable approximation for a weakly-interacting system,
and is also a necessary preparation for setting up thermal perturbation theory,
In the absence of interactions, the energies of the separate particles
are good quantum numbers.
To begin with, quantise the system in a finite volume $V$, so that the
single-boson energies $\epsilon_r$ are discrete.  We can choose as
the basis states $|i\rangle$ of
the system those labelled by the single-particle occupation numbers
$n_r$: 
$$
|n_1\,n_2\,n_3\dots\rangle
$$
and the eigenvalues of the noninteracting Hamiltonian $H_o$ are
$$
n_1 \epsilon_1 + n_2 \epsilon_2 + n_3 \epsilon_3 + \dots
$$
So the noninteracting grand partition function is 
$$
\eqalign{
Z_0 &= \sum_{\{n_r\}}  e^{-\beta (n_1 \epsilon_1 + n_2 \epsilon_2 +\dots )} \cr
&=  \prod_r \Big (\sum _{n_r}e^{-\beta n_r e_r}\Big ) = \prod_r {1\over
1-e^{-\beta \epsilon_r}}\cr} \eqno (6a) 
$$
and 
$$
\log Z_0 = - \sum_r \log \;(1 -e^{-\beta\epsilon_r}) \eqno (6b) 
$$
In the continuum limit
$$
\sum_r \rightarrow V \int {d^{3}k \over (2\pi)^3} \eqno (7) 
$$
and so the noninteracting equation of state is
$$
P = {T\over V} \log Z_0 = -T\int {d^3 k \over (2\pi )^3} \log (1
- e^{-\beta k^0)} \eqno (8)
$$
where $k^0 = \sqrt{{\bf k}^2 + m^2}$.  If the bosons have non zero
spin, there is an additional factor $g_s$ corresponding to the spin
degeneracy of each single-boson state.

In the continuum limit we usually work with fields
$$
\phi (x) = \int {d^{3}k\over (2\pi)^3}\; {1\over 2k^0} a({\bf k})e^{-ik\cdot
x} + \hbox{{  h.c.}} \eqno (9a)
$$
with
$$
[a ({\bf k}), a^{\dag} ({\bf k'})] = (2\pi)^3 2k^0 \delta^{(3)} ({\bf
k}-{\bf k'})\eqno (10a)
$$  
In the discrete case, we usually define
$$
[ a_r, a^{\dag}_s ] = \delta_{rs} \eqno (10b)
$$
If we sum this over $r$, the result is $1$.  But if we apply $V \int
d^3 k/(2\pi)^3$ to $(10a)$, the result is rather $2k^0 V$.  That is
(10a) and (10b) have definitions of the operators $a$ differing by a
factor $\sqrt{2k^0 V}$.  We correct for this by defining the field in
the discrete case to be
$$
\phi (x) = \sum_r {1\over \sqrt{2\epsilon_r V}} a_r e^{-i\epsilon _
r{t}} e^{i{\bf k}_r \cdot {\bf x}} + \hbox{{  h.c.}} \eqno (9b)
$$

The average number of particles in the single-particle level $\e _s$ is
$$\eqalignno{
\langle a^{\dag}_sa_s\rangle&=Z_0^{-1}\sum_{\{n_r\}}
\langle n_1\,n_2\,n_3\dots|e^{-\beta H_0}\,a^{\dag}_sa_s|n_1\,n_2\,n_3\dots\rangle
\cr
&=\left ({1\over 1-e^{-\beta\epsilon _s}}\right )^{-1}
\; \sum_{n_s} n _s \; e^{-n_s\beta\epsilon _s} =
f(\epsilon _s)&(11a)\cr}
$$
where $f$ is the Bose distribution
$$
f(\epsilon ) = {1\over e^{\beta\epsilon} -1} \eqno (11b)
$$
It is interesting to calculate the average of the square of the
number of particles in the single-particle level $\e _s$:
$$
\langle\big ( a^{\dag}_sa_s\big )^2\rangle
=\left ({1\over 1-e^{-\beta\epsilon _s}}\right )^{-1}
\; \sum_{n_s} n _s^2 \; e^{-n\beta\epsilon _s}=
\big (f(\e _s)\big )^2+f(\e _s)\big( f(\e _s)+1\big )
\eqno(11c)
$$
The average of the square is not equal to the square of the average, because
the number of particles in the level 
$\e _s$ fluctuates from system to system in the ensemble.

We can now calculate a noninteracting thermal Green's function,
defined as the thermal average
$$
\langle T \phi (x) \phi (0) \rangle_0 
= Z^{-1}_{0} \sum_i \langle i| e^{-\beta H_0} T
\phi (x) \phi (0) |i\rangle \eqno {(12)}
$$
Observe first that there is an operator identity
$$
T \phi (x) \phi (0) = \langle0| T \phi(x) \phi (0) |0\rangle + : \phi (x) \phi
(0): \eqno {(13)}
$$
where : : denotes the usual normal product. The first term contributes
to (12) just the usual zero-temperature Feynman propagator.  To
evaluate the contribution from the second, use the discrete case (9b)
and so obtain double sums $\textstyle{\sum_{r,s}}$ of terms 
$a^{\dag}_r a_s, \, a_r a_s,\,
a^{\dag}_r a^{\dag}_s$.  When we take the necessary expectation values, only
the first survives, and then only for $r=s$.  
So, using (11) we find that
$$
\langle: \phi (x) \phi (0):\rangle_0 = {1\over V} \sum_r {1\over 2\epsilon_r}
f(\epsilon_r) e^{i\epsilon_r t - i{\bf k}r \cdot {\bf x}} 
+ \hbox{{  h.c.}} \eqno
(14)
$$
Going to the continuum limit, we may write 
$$
\eqalign{
\langle T \phi (x) \phi (0) \rangle_0 &= \int {d^4 k\over (2\pi)^4}\;
e^{-ik\cdot x} D_T (k)\cr
D_T (k) = {i\over k^2 -m^2 + i\epsilon} &+ 2\pi \delta (k^2 - m^2)\,
n(k^0)\cr} \eqno (15) 
$$
where $n(k^0) = f(|k^{0}|)$.  The second term is the contribution from
the heat bath; it contains the $\delta$-function because so far the
heat-bath particles do not interact and so they are on shell.
\bigskip
{\bf Perturbation theory}

Suppose now that we introduce an interaction and again calculate
$\langle T \phi (x) \phi (0) \rangle$.  $\phi (x)$ is now the
interacting Heisenberg-picture field: it is the familiar operator of
zero-temperature field theory.  We can develop a perturbation theory
along the same lines as for zero-temperature scattering theory,
by introducing an
interaction picture that coincides with the Heisenberg picture at some
time $t_0$:
$$
\eqalign{
\phi_I (t, {\bf x}) &= \Lambda (t)\phi (t, {\bf x}) \Lambda^{-1} (t) \cr
\Lambda(t) &= e^{i (t - t_0)H_{0I} } e^{-i(t-t_0)H } \cr} \eqno (16)
$$
where $H_{0I}$ is the free-field Hamiltonian in the interaction
picture. Define as usual
$$
\eqalign{
\Lambda (t_1) \Lambda^{-1} (t_2) &= U(t_1, t_2) \cr}
$$
so that $\Lambda (t)=U(t,t_0)$. Because $U$ is constructed from ordinary
zero-temperature operators, as usual we have
$${U(t_1, t_2)
= T \exp \left ( -i \int^{t_1}_{t_2} dt\; H^{{\rm INT}}_I (t) \right )
} \eqno (17)
$$
We need $U(t_1, t_2)$ for complex $t_1$ and $t_2$, so that we
integrate $t$ along some contour $C$ running from $t_2$ to $t_1$ in the
complex plane and generalise the time ordering $T$ to an ordering
$T_c$ along $C$: the operator whose argument is nearest to $t_1$ along
the contour comes first.

Now, from the definition of $U$,
$$
e^{-\beta H} = e^{-\beta H_{0I}} \; U(t_0 -i\beta, t_0)
$$ 
so that
$$
\eqalign{
Z^{-1} {\rm tr} \; e^{-\beta H} \phi (x) \phi (0) &= 
 Z^{-1} {\rm tr}\; e^{-\beta H_{0I}} U(t_0 -i\beta , x^{0})\,\phi_{I}(x) 
U(x^0, 0)\, \phi_I (0)\, U (0, t_0) \cr
&= Z_0 Z^{-1}\big\langle U(t_0 - i\beta , x^0)\, \phi_I (x)\, U (x^{0}, 0)
\phi _I(0)\, U(0, t_{0}) \big\rangle_0 \cr}\eqno (18) 
$$
where I have used $U(t_1,t_2)U(t_2,t_3)=U(t_1,t_3)$.

Compare with what one has at zero temperature:
$$
\langle 0 | \phi (x) \phi (0) |0\rangle = \langle 0 | U (\infty,
x^0) \phi_I (x) U (x^{0},0 ) \phi_I (0) U(0,-\infty ) |0\rangle
$$ 
In (18) we have a non-interacting thermal average instead of a vacuum
expectation value, the first argument of the first $U$ is
$(t_0 -i\beta )$ instead of $\infty$, and the second argument of the last
$U$ is $t_0$
instead of $-\infty$.  In fact there is a close similarity between
thermal perturbation theory and the usual zero-temperature Feynman
perturbation theory. One draws familiar-looking Feynman graphs, now called
thermal graphs,
and the main difference is that instead of the
internal lines in the ordinary Feynman graphs representing vacuum expectation
values, in the thermal graphs 
they are non-interacting thermal averages.  

In order to derive
this, one needs to establish Wick's theorem.  It is a remarkable fact
that indeed, for example,
$$
\langle T_c \phi_I (x_1) \phi_I (x_2) \phi_I (x_3) \phi_I (x_4)
\rangle_{0} = \sum \langle T\phi_I \phi_I \rangle_0\, \langle T
\phi_I \phi_I \rangle_0 \eqno (19a)
$$
where the sum is over the possible pairings of the fields.  For almost
every ensemble other than one in thermal equilibrium there would be
correction terms to (19a). To understand this, go back to the discrete
case. One of the relations needed to establish (19a) is
$$
\langle a^{\dag}_ra^{\dag}_sa_ta_u\rangle=
\langle a^{\dag}_ra_t\rangle\langle a^{\dag}_sa_u\rangle
+\langle a^{\dag}_ra_u\rangle\langle a^{\dag}_sa_t\rangle
\eqno(19b)
$$
This relation is straightforward except when the indices
$r,s,t,u$ are all equal. Then, from (11), the right-hand side is just
$2\big (f(\e _r)\big )^2$, while with (10b) the left-hand side is
$$
\langle (a^{\dag}_ra_r)^2-(a^{\dag}_ra_r)\rangle
$$
which we calculate using (11c). We get the same result as for the
right-hand side because of the fluctuation term $f(f+1)$ in (11c);
for most other ensembles this term would be different
and so Wick's theorem would not be valid.

One needs to choose a value for $t_0$.  A common choice is $t_0 = 0$,
with the contour $C$ for the $t$ integrations running from $0$ to
$-i\beta$ along the imaginary axis.  This is the imaginary-time
formalism.  Alternatively, $t_0 \rightarrow -\infty$,  which with suitable
contour choice gives the real-time formalism. Although under certain
circumstances the imaginary-time formalism is a powerful calculational
tool, the real-time formalism stays much closer to the physics and is more 
versatile.
\bigskip
{\bf Real-time formalism}

One is interested in equilibrium properties of the plasma at finite
times.  Presumably these are independent of how it reached thermal
equilibrium.  So, as in familiar scattering theory, we are free to
imagine that the interaction slowly switches off as we go into the
remote past, and then when we take $t_0\rightarrow -\infty$ the
interaction-picture fields become the usual noninteracting {\sl in}
fields.  The corresponding {\sl in} states are direct products of
non-interacting single-particle states.  We need to choose how the
contour $C$ runs from $-\infty$ to   $(-\infty -i\beta )$, and the
choice that keeps the formalism in the most direct contact with
the physics is the  so-called Keldysh
one: along the real axis from $-\infty$ to $\infty$, back to
$-\infty$, then straight down to $(-\infty -i\beta )$:
$$
\matrix{\hbox{\epsfxsize=80truemm  \epsfbox{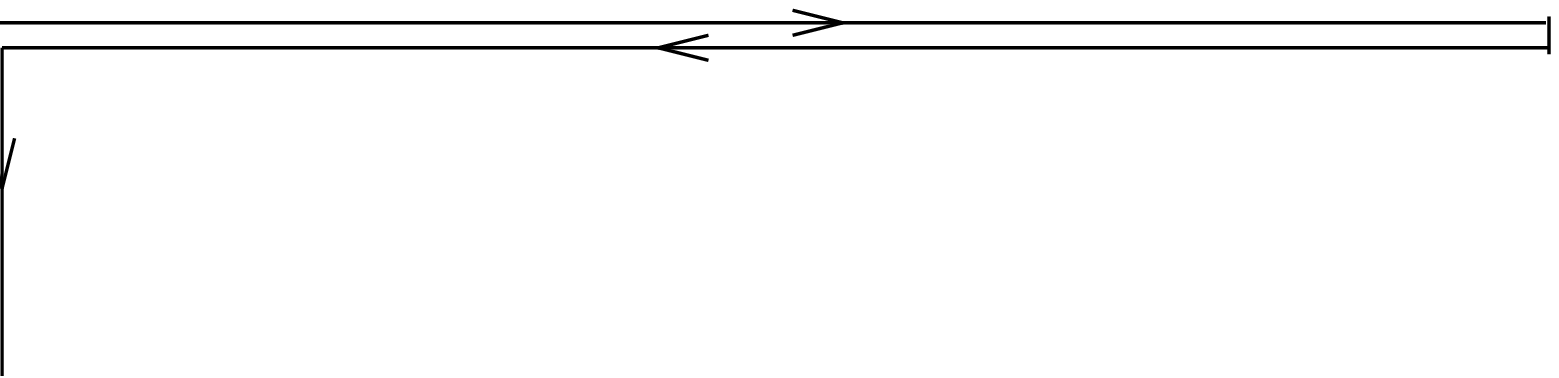}}\cr}
$$

For most applications (though not all\defref\evans{
T S Evans and A C Pearson, Physical Review D52 (1995) 4652
}) it turns out that the
vertical part of the contour may be omitted.  Then we may write
the propagator that corresponds to a line of a thermal
graph as a $2\times 2$ matrix:
$$\eqalign{
{\bf D} (x_1, x_2) &= \left [\matrix{\langle T \phi_{{\rm in}} (x_1)
\phi_{{\rm in}} (x_2) \rangle_0 & \langle \phi_{{\rm in}} (x_2)
\phi_{{\rm in}} (x_1) \rangle_0 \cr
\langle\phi_{\rm_{in}} (x_1) \phi_{{\rm in}} (x_2) \rangle_0 &
\langle\bar{T} \phi_{{\rm in}}(x_1)\phi_{{\rm in}} (x_2) \rangle_0 \cr}\right ]}
\eqno(20)
$$

When both $x_1$ and $x_2$ are on the $-\infty$ to $\infty$ part of
$C$, the ordering $T_c$ along the contour is ordinary time ordering $T$; this
corresponds to the element $D_{11}$ of $\bf{D}$.  When both are on
the $\infty$ to $-\infty$ part of $C$, $T_c$ is anti-time-ordering
$\bar{T}$; this corresponds to $D_{22}$.  The off-diagonal elements
correspond to $x_1$ being on one part of $C$ and $x_2$ on the other.
There is translation invariance: the elements of ${\bf D}$ depend only on
the difference between $x_1$ and $x_2$.

I have already shown how to calculate $D_{11}$; the result is given
in (15).  The other elements of $\bf{D}$ may be calculated in the
same way, and its Fourier transform is
$$
{\bf D}(k)=\left[\matrix{i\over {k^2-m^2 + i\epsilon} &2\pi\delta ^-(k^2-m^2)\cr
2\pi\delta^+(k^2-m^2)&
{-i\over k^2-m^2-i\epsilon}\cr} \right ]+ 2\pi \delta (k^2 - m^2)\,n(k^0)  
\left[\matrix{ 
1 & 1 \cr 1 & 1 \cr}\right]\eqno (21a)
$$
It may also be written in the form
$$
{\bf D}(k)={\bf M}(k^0) \ \tilde{\bf D}(k) \  {\bf M}(k^0)\eqno (21b)
$$
with
$$\eqalignno{
{\bf M}(k^0) = & \sqrt{n(k^0)} \left[\matrix{ e^{{1\over 2} \beta |k^0|} & 
e^{-{1\over 2}\beta k^0} \cr e^{{1\over 2}\beta k^0} & e^{{1\over 
2}\beta |k^0|}\cr}\right ]\cr
\tilde{\bf D}(k) = &  \left[ \matrix{D(k)&0\cr
                   0&  D^*(k)\cr}\right ]\cr
&D(k)={i\over {k^2-m^2 + i\epsilon}} &(21c)\cr}
$$
For the case of a fermion field, there is a rather similar matrix
propagator, but with the Fermi-Dirac distribution replacing the Bose
distribution.

\bigskip
\goodbreak
{\bf Matrix structure}

The elements of the matrix propagator (20) are not independent. For example,
$$
\eqalignno{
D_{21} (x) &= \langle \phi_{{\rm in}} (x) \phi_{{\rm in}} (0)
\rangle_0 = Z_0^{-1} {\rm tr}\; e^{-\beta H_{0{\rm in}}} \phi_{{\rm in}}
(x) \phi_{\rm in} (0) \cr
&= Z_0^{-1} {\rm tr}\; e^{-\beta H_{0{\rm in}}} \phi_{{\rm in}}(0)
e^{\beta H_{0{\rm in}}} \phi_{{\rm in}}(x) e^{{\beta H_0{\rm in}}} \cr
&= Z_0^{-1} {\rm tr}\; e^{-\beta H_{0{\rm in}}} \phi (0) \,\phi (x^0 - i\beta
, {\bf x}) \cr
&=D_{12} (x^0 - i\beta , {\bf x})&(22a)\cr}
$$
Here, I have used a general property of traces, that ${\rm tr}(AB) =
{\rm tr} (BA)$, and the fact that $H_{0{\rm in}}$ is the time-translation
operator for the noninteracting field $\phi_{{\rm in}}$.  
The Fourier transform of (22a) is
$$
D_{12} (k) = e^{-\beta k^0} D_{21} ({k}) \eqno (22b)
$$
Also, from their definitions (20), one can see that $D_{11}$ and $D_{22}$
may be expressed in terms of $D_{12}$ and $D_{21}$.  
It is this, together with (22b),
which is responsible for the matrix structure (21b).  

Define
a dressed thermal propagator matrix  $ {{\bf D}}' (x_{1}, x_{2})$
analogous to ${\bf D} (x_1, x_2)$ in (20), but with the
interacting Heisenberg field instead of $\phi_{{\rm in}}$.  
For example, 
$$
D'_{12} (x_1, x_2)=
\langle \phi (x_2) \phi (x_1)\rangle
\eqno(22c)
$$  
Then, because $H$ is the
time-translation operator for $\phi$, we can again derive
$$
D'_{12} (k) = e^{-\beta k^0} D'_{21} (k) \eqno (22d)
$$
and so deduce that  ${\bf D}'$ has a matrix structure similar to
that of ${\bf D}$:
$$
{\bf D}'(k) = {\bf M} \pmatrix{D'(k) & 0 \cr
0 & D^{\prime *}(k)} {\bf M} 
\eqno(23)
$$
where ${\bf M}(k^0)$ is the same matrix as in (21c).
Define the thermal self-energy matrix ${\bf \Pi}$ by
$$
-i{\bf \Pi} = {\bf D}^{-1} - {\bf D'}^{-1} \eqno (24a)
$$
Then it follows that ${\bf\Pi}$ has the structure 
$$
-i{\bf \Pi} = {\bf M}^{-1} \pmatrix{-i\Pi(k,T) & 0 \cr
0 & [-i\Pi(k, T)]^*\cr}{\bf M}^{-1} \eqno (24b)
$$
If we then solve (24a) for ${\bf D'}$, we find
$$
{\bf D}' = {\bf M} \pmatrix {i\over k^2 - m^2 - \Pi & 0 \cr
0 & {-i \over k^2 - m^2 - \Pi^{*}} \cr} {\bf M} \eqno (25)
$$
So it is natural to interpret ${\rm Re}\;\Pi$ as a temperature-%
dependent shift to the mass $m^2$.  $\Pi$ also has an imaginary part, so
the propagation of the field through the heat bath decays with time.

In scalar field theory,
$$
-i{\bf\Pi}=\matrix{\hbox{\epsfxsize=70truemm  \epsfbox{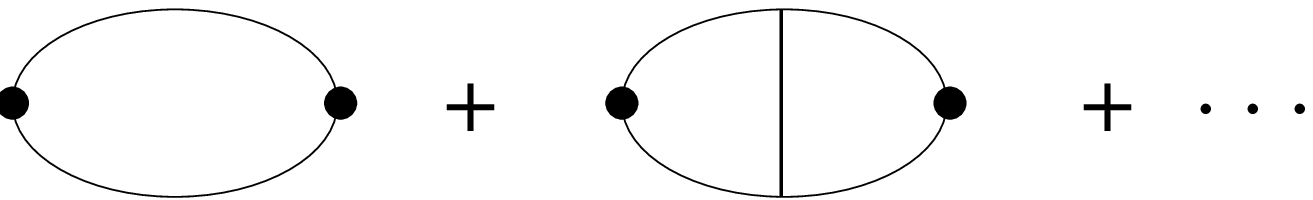}}\cr}
$$
To calculate the contribution to $\Pi_{12}$ from the second term, for
example, one needs
$$
\matrix{\hbox{\epsfxsize=120truemm  \epsfbox{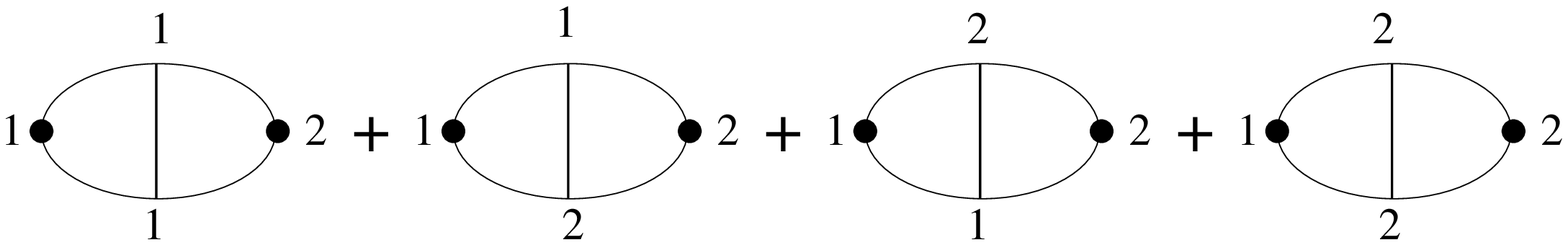}}\cr}
$$
where each line $\matrix{\hbox{\epsfxsize=13truemm  \epsfbox{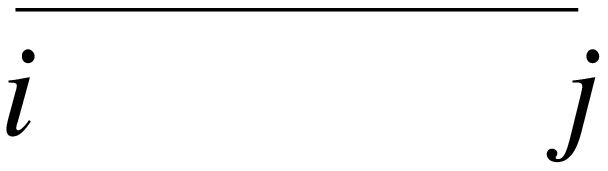}}\cr}$
 represents $D_{ij} (k)$ and each vertex $2$ is
the same as the normal vertex $1$, but opposite in sign.
\bigskip
{\bf Imaginary-time formalism}

The real-time formalism stays close to the physics, but has the calculational
complication that the propagator is a matrix. In the imaginary-time formalism
there is not this complication, though except for a few simple cases
there is  the need to perform an anlytic continuation from imaginary
to real time at the end of the calculation.
In the imaginary-time formalism  
the $t$ integration runs along the imaginary axis, so ordinary
time-ordering is replaced with ordering in imaginary time:
$$
\bar{D} (x_1, x_2) = \theta (-{\rm Im}\; t) D_{21} (x_1, x_2) + \theta
({\rm Im}\; t)
D_{12} (x_{1}, x_{2}) \eqno (26) 
$$
where $t = x_1^0 - x^0_2$.  Because both $x^0_1$ and $x^0_2$ are
integrated from $0$ to $i\beta$, we need $\bar{D} (x_{1},x_{2})$ for
values of ${\rm Im}\; t$ in the range $-\beta$ to $+\beta$.  In this finite
interval it has a Fourier-series expansion:
$$
\bar{D} (t, {\bf x}) = {i\over \beta} \sum^\infty_{n=-\infty} D_n
({\bf x}) \, e^{\omega _n t} \eqno (27a)
$$
where $\omega _n = n\pi/\beta$.  However, the relation (22a) implies that
$\bar{D} (t,x) = \bar{D} (t+i\beta, {\bf x})$, so that only even
values of $n$ contribute to the sum.  In the case of fermions, the
anticommutativity of the fields results in a minus sign appearing in the
corresponding relation (22a), and so then only odd values of $n$
contribute.

If we apply a $3$-dimensional Fourier transformation to (27a) and
invert the Fourier summation over $n$, we find
$$
D_n ({\bf k}) = \int^{i\beta}_{0} dt\; e^{-w_n t} D_{21} (t, {\bf k})
\eqno (27b)
$$
which turns out to be just the ordinary Feynman propagator with $k^0 =
i\omega _n$.  So the Feynman rules are just like the zero-temperature ones,
except that the energy-conserving $\delta$-function at each vertex is
replaced with a Konecker delta which imposes conservation of the
discrete energy, and round each loop of a thermal graph
$$
\int {d^4 k\over (2\pi)^4} \rightarrow {i\over\beta} {\sum_n} \int
{d^{3}k\over (2\pi)^3} \eqno (28)
$$

{\bf Gauge theories}

For gauge theories there is the complication that the grand partition
function has to include the projection operator ${\Bbb P}$ onto
physical states: see (3).  The manipulations that set up the scalar-field
perturbation theory include the use of the basic property of traces,
trace $AB=$trace $BA$. To follow exactly the same route for a gauge theory one
would need to use trace $AB{\Bbb P}=$ trace $BA{\Bbb P}$, but this
is not a valid identity. It turns out\defref\rebhan{
P V Landshoff and A Rebhan, Nuclear Physics B383 (1992) 607 and B410 (1993) 23
}
that there are two formalisms for perturbation theory for a thermal
gauge theory:
{\parindent=8truemm
\item{1} Only the two physical degrees of freedom of the gauge field (the
transverse polarisations) acquire the additional thermal propagator;
the other components of the gauge field, and the ghosts, remain frozen
at zero temperature.  (This is for the bare propagators; self-energy
insertions in the unphysical bare propagators do depend on the
temperature.)
\item{2} All components of the gauge field, and the ghosts, become heated to
temperature $T$.}

In the zero-temperature field theory, the ghosts are introduced in
order to cancel unwanted contributions from the unphysical components
of the gauge field\defref\three{
T Kugo and I Ojima, Physics Letters 73B (1978) 459
}, and the two formalisms of the thermal perturbation
theory lead to the same answers
for calculations of physical quantities for that reason.  Often, using
formalism $1$ makes calculations simpler.  It also makes them stay
closer to the physics.
\bigskip
{\bf Photon or dilepton emission from a plasma}

As an application, consider the emission of a real or virtual photon
of momentum $q$ from a quark-gluon plasma.  This is supposed to be an
important diagnostic test of whether a plasma has been created in an
experiment and has reached thermal equilibrium, and is then
a way to measure its
temperature.

Before a photon is emitted, the plasma is described by the density
matrix
$$
\rho = Z^{-1} \sum_i | i\, {\rm in} \rangle\langle i\, {\rm in} | e^{-\beta H} \eqno
(29)
$$
The emission probability is calculated from squared matrix elements of
the Heisenberg-picture electromagnetic current\defref\mclerran{
L D McLerran and T Toimela, Physical Review D31 (1985) 545\h
H A Weldon, Physical Review D42 (1990) 2384
}:
$$
W^{\mu\nu} (q) = Z^{-1} \sum_{f} \int d^4 x e^{iq\cdot x} \langle f\,
{\rm out} | J^{\mu} (x)\left ( \sum_{i}| i\, {\rm in}\rangle\langle
 i\, {\rm in}| e^{-\beta H} \right )J^{\nu} (0) | f\, {\rm
out\rangle } \eqno (30a)
$$  
where I have introduced also a complete set of out states for the
plasma.  These satisfy the completeness relation
$$
\sum_f | f{\rm out} \rangle\langle f {\rm out} | = 1 
$$
and so
$$
W^{\mu\nu} (q) = Z^{-1} \sum_i \int d^4 x\, e^{iq\cdot x} \langle
i\, {\rm in}|e^{-\beta H} J^{\nu} (0) J^{\mu} (x) |i\, {\rm in}\rangle 
\eqno (30b)
$$

If we introduce a matrix ${\bf G}^{\mu\nu}(q)$ analagous to ${\bf D}'$, but
with thermal averages of products of electromagnetic currents  instead
of fields, $W^{\mu\nu} (q)$ is just $G^{\mu\nu}_{12} (q)$.  So we draw
thermal graphs where current $q$ enters at a $1$ vertex and leaves
at a $2$ vertex, and distribute the labels $1$ and $2$ in all possible
ways on the other vertices.  For example,
$$
\matrix{\hbox{\epsfxsize=120truemm  \epsfbox{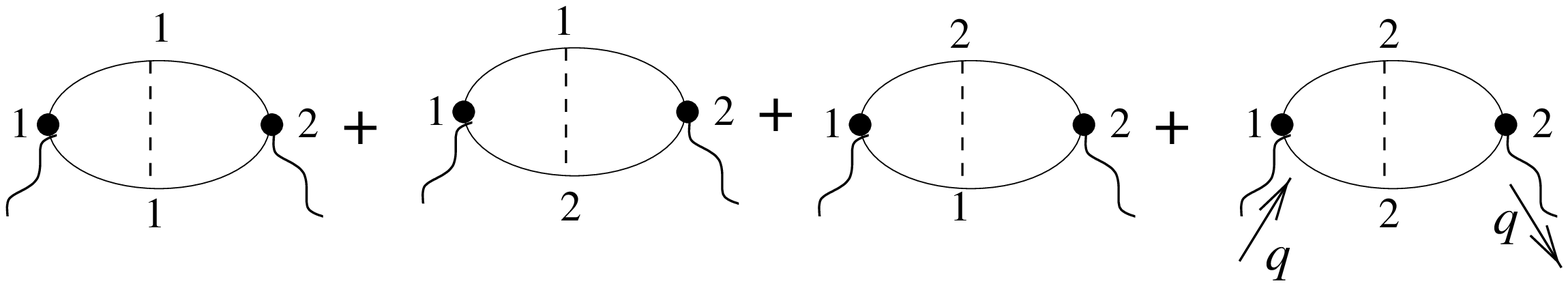}}\cr}~~~~~\hbox{{\sevenrm
(thermal graphs)}}
$$
where the solid lines are quarks and the dashed lines are gluons.
The emission rate is calculated from a sum of matrix elements times their
complex
conjugates; the vertices labelled 1 correspond to a contribution to
the matrix element and those labelled 2 to its complex conjugate.
The sets of 1-vertices and of 2-vertices are joined by (12)-lines
which, according to (21a) are on shell and represent particles in the heat bath.
According to (21a),
$$
D_{12}(k)=\left\{\matrix{2\pi\delta(k^2-m^2)\,n(k^0)&(k^0>0)\cr
                      2\pi\delta(k^2-m^2)\,(1+n(k^0))&(k^0<0)\cr}\right .
\eqno(31)
$$
So, because the density of particles with momentum $k$ in the heat bath is
$n(k^0)$,  when $k^0>0$ a  (12)-line represents a particle with 
momentum entering the photon-production reaction from the heat bath, while when
$k^0<0$ it represents instead stimulated emission of a particle with
momentum $-k$ into the heat bath. In the loop integrations, the 
energy on each line extends over both positive and negative values,
so the thermal graphs sum together many physical processes.  Consider the
first graph, for example. Its right-hand part, with the 2-vertex,
represents the processes
$$
\matrix{\hbox{\epsfxsize=85truemm  \epsfbox{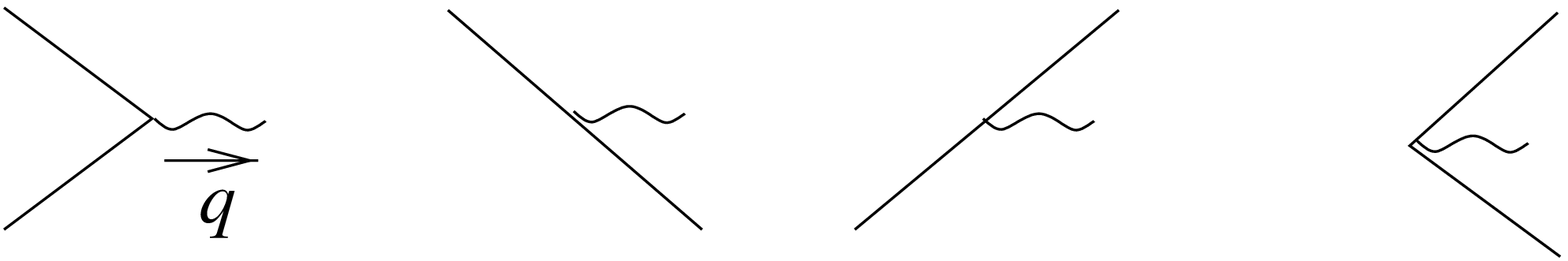}}\cr}~~~~~~~~~\hbox{{\sevenrm
(ordinary Feynman graphs)}}
$$
depending on the signs of the relevant internal energies.
In fact, energy-momentum conservation allows only the first one to be
non-zero.  I have not drawn in the other heat-bath particles, but
remember that they are there as spectators.  The other part of the thermal
graph contains 1-vertices and  (11)-lines, and according to (21a) the
(11)-propagator is the ordinary Feynman zero-temperature propagator, plus
a thermal part.  If I use only the zero-temperature
part of $D_{11}$ in each of these lines, I obtain the amplitude
$$
\matrix{\hbox{\epsfxsize=20truemm  \epsfbox{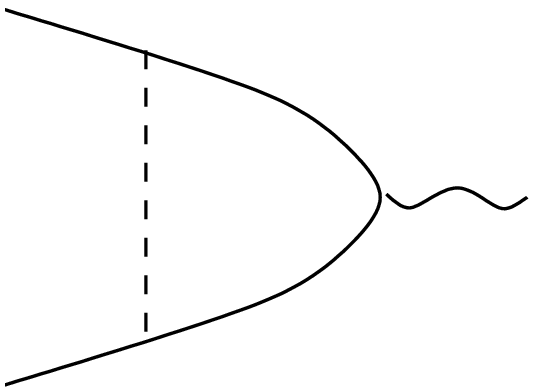}}\cr}~~~~~~~~~\hbox{{\sevenrm
(ordinary Feynman graph)}}
$$
(plus other terms which again vanish for kinematic reasons)
and so part of the thermal graph represents the interference between this and
$\matrix{\hbox{\epsfxsize=8truemm  \epsfbox{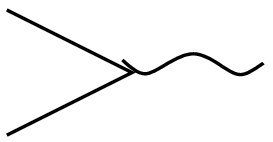}}\cr}$.  If instead I use 
the thermal part $n(k^0)\,2\pi\delta(k^2)$  of the (11) gluon
propagator, I obtain the amplitudes
$$
\matrix{\hbox{\epsfxsize=65truemm  \epsfbox{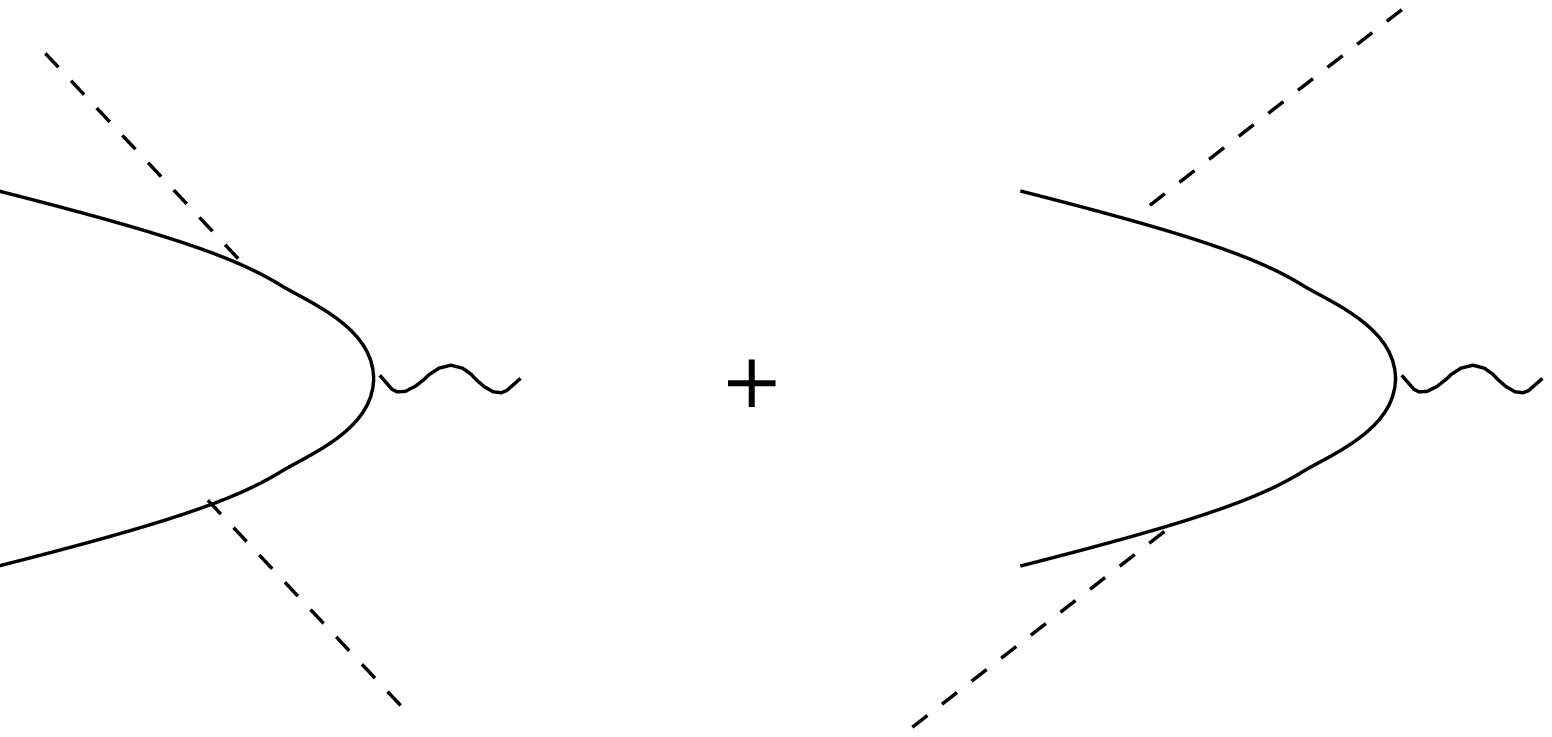}}\cr}~~~~~~~~~\hbox{{\sevenrm
(ordinary Feynman graphs)}}
$$
In each case, the incoming and outgoing gluon lines must have the same
momentum $k$, so that these amplitudes again interefere with 
$\matrix{\hbox{\epsfxsize=8truemm  \epsfbox{5a.eps}}\cr}$,
but now with the gluon $k$ being one of the spectator particles in
the heat bath: 

$$
\matrix{\hbox{\epsfxsize=25truemm  \epsfbox{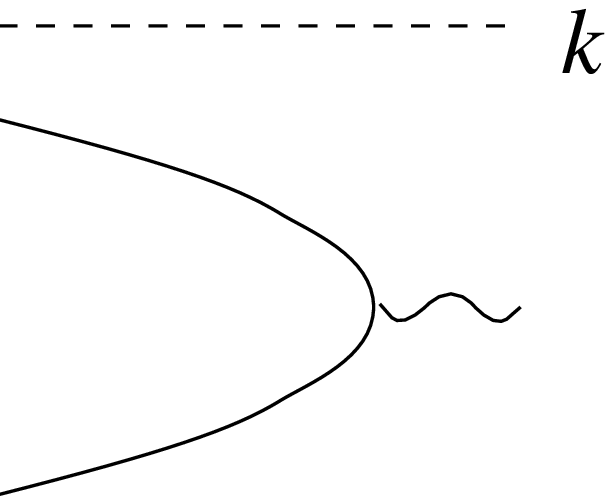}}\cr}~~~~~~~~~\hbox{{\sevenrm
(ordinary Feynman graph)}}
$$

Similarly, I  can identify physical processes that involve the thermal
parts of the (11) quark propagators.

Even a simple-looking thermal graph corresponds to a large number of
physical processes\defref\niegawa{
N Ashida et al, Physical Review D45 (1992) 2066\h
P V Landshoff, Physics Letters B386 (1996) 291
}, each of which can be rather complicated.  An
example is
$$
\matrix{\hbox{\epsfxsize=45truemm  \epsfbox{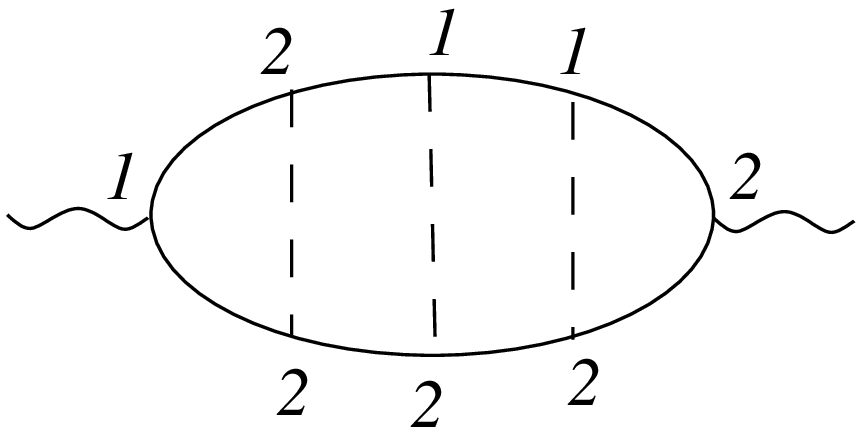}}\cr}~~~~~~~~~\hbox{{\sevenrm
(thermal graph)}}
$$
for which just one of the physical processes is the interference between
$$
\matrix{\hbox{\epsfxsize=65truemm  \epsfbox{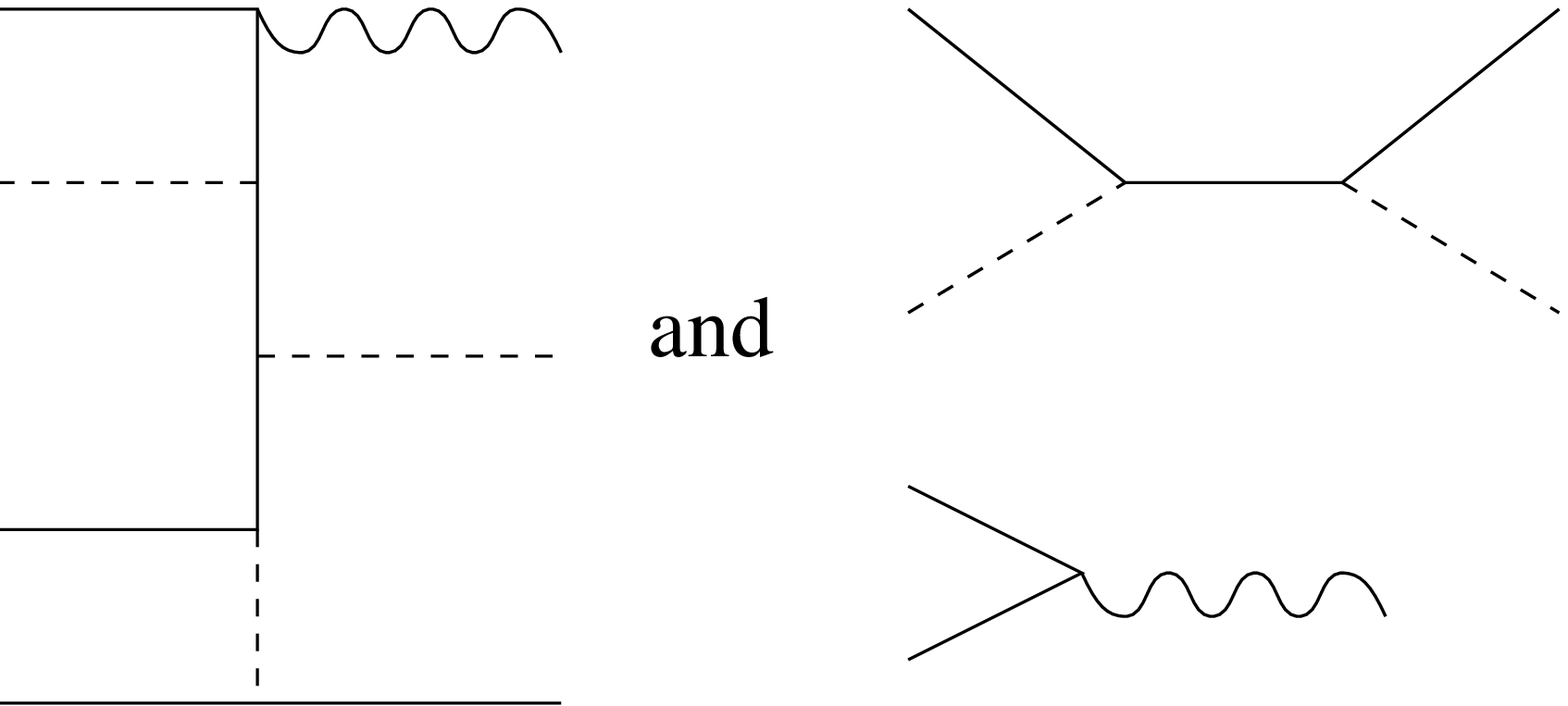}}\cr}~~~~~~~~~\hbox{{\sevenrm
(ordinary Feynman graphs)}}
$$  
A disconnected Feynman graph occurs because the thermal graph has 
``islands'' of groups of {1}-vertices entirely surrounded by
{2}-vertices.
\bigskip
{\bf Resummation and discontinuities}

The matrix amplitude ${\bf G}^{\mu\nu}(q,T)$ has structure similar to that
of ${\bf D}'$ in (23):
$$
{\bf G}(q,T)={\bf M}\left (
    \matrix{ G(q,T)&0\cr
             0&G^{*}(q,T)\cr}\right ){\bf M}
\eqno(32)
$$
From the explicit form (21c) of ${\bf M}$ we find that
$$\eqalignno{
G_{12}(q,T)&=e^{\half\beta(|q^0|-q^0)}n(q^0) \Big(G(q,T)+G^{*}(q,T)\Big)\cr
G_{11}(q,T)&=n(q^0)\left (e^{\beta q^0}G(q,T)+G^{*}(q,T)\right )&(33)\cr}
$$
So, for positive $q^0$, 
$$
G_{12}(q,T)=2n(q^0)\hbox {Re }G(q,T)={2\over e^{\beta q^0}+1}\hbox{ Re }
G_{11}(q,T)
\eqno(34a)
$$
With a small change of notation, ${\bf G}(q,T)=i{\bf F}(q,T)$, 
this becomes
$$
-iF_{12}(q,T)=2n(q^0)\hbox{ Im }F(q,T)=
{2\over e^{\beta q^0}+1}\hbox{ Im }F_{11}(q,T)
\eqno(34b)
$$
This relation between the (12) and (11) elements of ${\bf F}(q,T)$ is
reminiscent of the zero-temperature unitarity relation, which expresses
the imaginary part of the forward elastic scattering amplitude
(which is the Fourier transform of a matrix element of a time-ordered
product of fields) to the 
sum of integrated squared amplitudes needed to calculate reaction 
rates\defref\elop{
R J Eden, P V Landshoff, D I Olive and J C Polkinghorne, {\sl
The analytic S-Matrix}, Cambridge University Press (1966)
}. So here the imaginary part of $F_{11}(q,T)$ (which 
is the Fourier transform of a thermal average of a time-ordered product of
currents) is related to $F_{12}(q,T)$, which up to a factor of
$i$ is the sum (30a)
of integrated squared amplitudes needed to calculate a thermal reaction rate.

At zero temperature the amplitude is real for $q^2<4m^2$, and
for $q^2>4m^2$ twice its imaginary part 
is the discontinuity across the set of branch cuts
that run along the positive real
axis from $q^2=4m^2$ to infinity.
Such a relationship
holds for each Feynman graph that contributes to the amplitude; the
Feynman-graph relationship is known as the cutting rule, and is named after
Cutkosky who first discovered it\ref{\elop}.

There are also cutting rules at nonzero temperature\ref{\lebellac}.
The simplest graph that contributes to ${\bf F}(q,T)$ is
$$
\matrix{\hbox{\epsfxsize=35truemm  \epsfbox{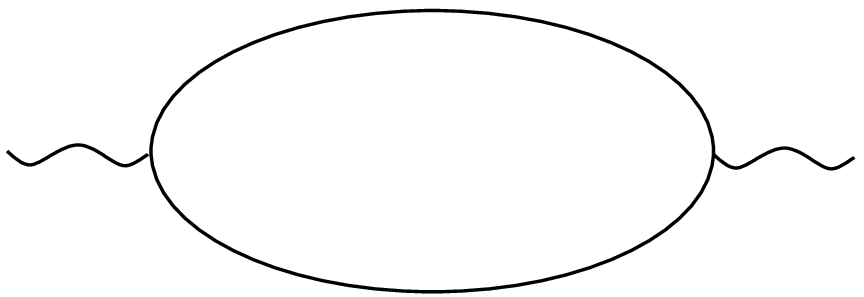}}\cr}~~~~~~~~~\hbox{{\sevenrm
(thermal graph)}}
$$
This graph has a branch cut that runs from $q^2=4m^2$ to $\infty$, like at zero temperature,
and another from $q^2=0$ to $-\infty$. It is real between these cuts.
The discontinuities across each of these are
proportional to the imaginary part of $F_{11}(q,T)$. However, this fact is
not very useful, because for higher-order graphs there is no gap between
the branch cuts and no region where $F_{11}(q,T)$ is real: it cannot be
analytically continued from the upper side of the cuts to the lower side, 
and so the notion of a discontinuity is not useful.
The relation (34b) between
$F_{12}(q,T)$, from which the reaction rate is calculated, and the imaginary
part of $F_{11}(q,T)$ is still valid, but it is no longer the discontinuity
across a cut.

When we calculate $F_{11}(q,T)$,
the internal lines of the graph correspond to thermal propagators
$D_{11}(k)$ and $D_{11}(q-k)$. From the explicit form (21a) of $D_{11}$, 
we see that these internal propagators are singular at $k^2=m^2$ and
$(q-k)^2=m^2$. When we perform the necessary integration over the
internal loop momentum $k$, these singularities generate\ref{\elop}
the branch points at $q^2=0$ and $4m^2$. However, Dyson resummation
replaces the bare propagators $D_{11}$ with dressed propagators
$D'_{11}$,
$$
\matrix{\hbox{\epsfxsize=160truemm  \epsfbox{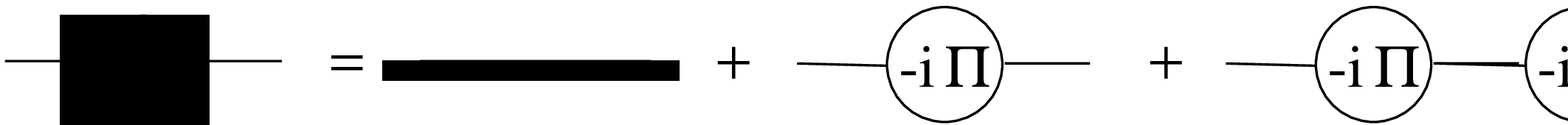}}\cr}
$$
so that the graph becomes
$$
\matrix{\hbox{\epsfxsize=40truemm  \epsfbox{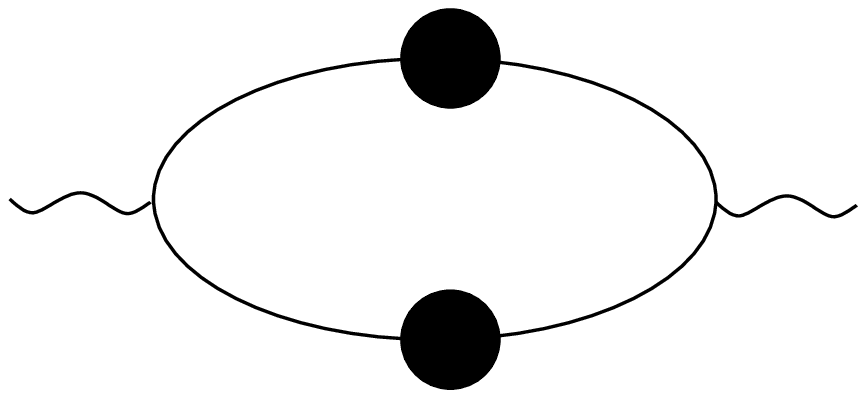}}\cr}~~~~~~~~~\hbox{{\sevenrm
(thermal graph)}}
$$
The resummation removes
the singularity of $D_{11}(k)$ at $k^2=m^2$ and replaces it with a
singularity whose position, according to (25), is
given by
$$
k^2=m^2+\Pi(k,T)~~~~~~~~~~~ \hbox{ or }~~~~~~~~~~~ k^2=m^2+\Pi^*(k,T)
\eqno(35)
$$
The consequence is that the branch point of $F_{11}$ that
was at $q^2=4m^2$ before the resummation has now moved off the real $q^2$ 
axis\defref\weldon{
H A Weldon, hep-ph/9806325
}.
But along the real axis the relation (34b) between $F_{12}(q,T)$ 
and the imaginary part of $F_{11}(q,T)$ is still valid.
\bigskip
{\bf Infrared divergences}

The infrared divergences of zero-temperature become much worse at
finite temperature: the Bose distribution diverges at zero energy and
causes the usual logarithmic divergences to become power divergences.
We know that the infrared divergences must cancel if the theory is to
make sense, and in practice they always do, but there is no general
theory yet to show this.  To some extent, the situation can be rescued by
including thermal self-energy insertions in the propagators, so that
they acquire a mass proportional to the temperature.  But, in the case
of photons or gluons, not all the degrees of freedom have a mass,
according to perturbation theory.  If the so-called magnetic mass is
non zero, it is nonperturbative.  

Consider, for example, the effect on
the decay rate $\pi^0 \rightarrow e^+ e^-$ of the microwave
background, which is a heat bath consisting only of photons\defref\jacob{
M Jacob and P V Landshoff, Physics Letters B281 (1992) 114
}.  Let
$\Gamma$ be the decay rate in vacuum.  The heat bath will change it
partly because it gives the electrons an additional
temperature-dependent mass $\delta m^2_e \propto e^2 T^2$.  This causes
a change $\delta\Gamma = \delta m_e^2 \,\partial\Gamma/\partial m^2_e$,
which is associated with thermal graphs of the form
$$
\matrix{\hbox{\epsfxsize=35truemm  \epsfbox{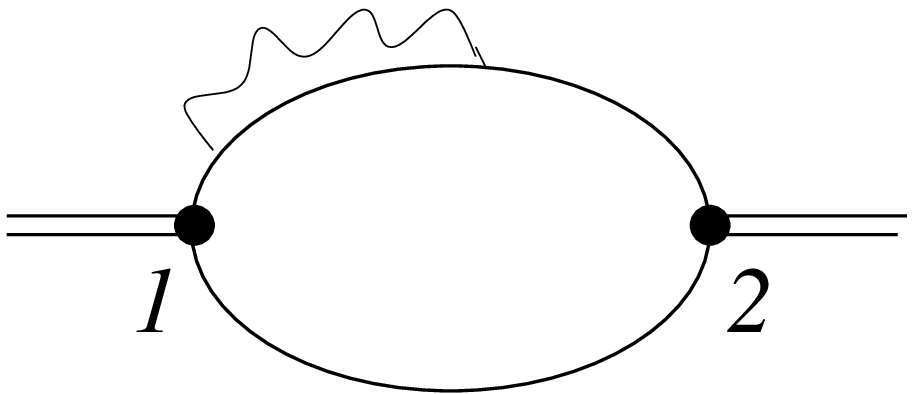}}\cr}
$$
There is also the thermal graph
$$
\matrix{\hbox{\epsfxsize=35truemm  \epsfbox{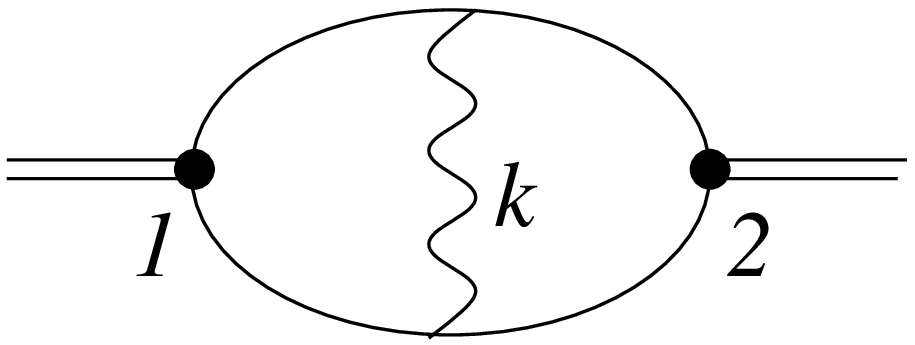}}\cr}
$$
One finds that 
$$
\eqalign{
{\Delta\Gamma\over\Gamma} ={\delta m_e^2\over\Gamma} \,{\partial\Gamma
\over\partial m^2_e}+
& {m_{\pi}\alpha _{EM} \over \pi ^3Q}
 \int d^4 k \,\delta (k^{2}) n(k^0)
\int d^4 p_1 d^4 p_2\, \delta^{(+)} (p_{1}^{2} - m^2_e )
\delta^{(+)} (p^2 - m_e^2 ) \cr
&\qquad\left ( {p_1\over p_1 \cdot k} - {p \over p_2 \cdot k}\right )^2
\big\{ \delta^{(4)} (p_1 + p_2 - P) - \delta^{(4)} (p_1 + p_2 + k - P)
\big\}\cr} \eqno {(36)}
$$
where $Q^2 = ({1\over 4} m^2_{\pi} - m^2_e )$.
In the integral, the first $\delta^{(4)}$-function corresponds to the
contribution from the internal vertices in the thermal graphs
 being both 1
or both 2
and the second to  12 and  21.
Each term separately is infrared
divergent, like $\int dk/k^2$, but the divergences  cancel.

In fact there is more cancellation than just that of the infrared
divergences.  For $T \ll Q$ one can expand (36) in powers of $T^2/Q^2$.  
One finds that
the first term, of order $\alpha _{EM} T^2/Q^2$, exactly cancels the
electron-mass-shift contribution $\delta \Gamma$, so that the net
change in the decay rate is of order $\alpha _{EM} T^4$.  The lesson is that,
in thermal field theory, it is of importance to calculate all terms;
gauge theories at finite temperature are rife with cancellations.

Another example where there is an apparent infrared problem is that of
the calculation of the equation of state for the quark-gluon plasma.
For instance, in the purely gluonic case thermal graphs of the form
$$
\matrix{\hbox{\epsfxsize=50truemm  \epsfbox{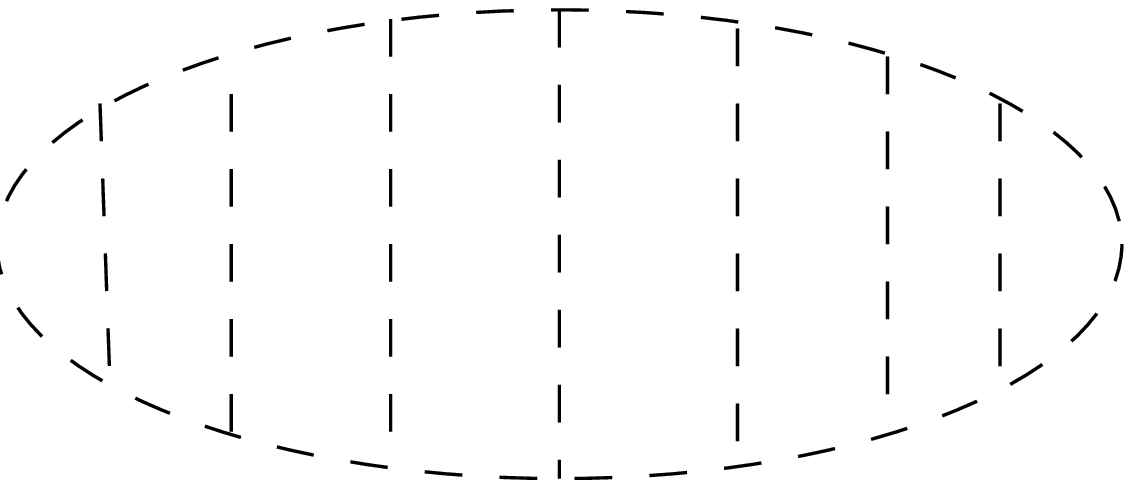}}\cr}
$$
become more and more divergent as more vertical lines are added.  
However, the problem goes away if one sums
over all thermal graphs\defref\horgan{
I T Drummond, R R Horgan, P V Landshoff and A Rebhan, Physics Letters
B398 (1997) 326}. 
Consider scalar field theory, where the 
Hamiltonian $H$ contains a term $\int d^3x (-\half m^2\phi ^2)$.  
Insert this into $Z=$ tr $e^{-H/T}$ and differentiate with respect to
$m^2$: 
$$
T{\pd\over\pd m^2}\log Z=-{1\over 2} \int d^3x\langle (\phi(x)) ^2\rangle
=\langle (\phi(0)) ^2\rangle V
\eqno(37a)
$$
Here I have used translation invariance:
$\langle (\phi(x)) ^2\rangle$ is independent of $x$. 
With the relation \hbox{$PV=T\log Z$}
from (4) and the definition (22c) of $D'_{12}(x)$,
this gives
$$
{\pd P\over\pd m^2}=-{1\over 2} D'_{12}(0)
=-{1\over 2}\int {d^4q\over (2\pi)^4}D'_{12}(q)
\eqno(37b)
$$
Insert the form of $D'_{12}(q)$ obtained from the matrix product
${\bf D}'={\bf M \tilde D}'{\bf M}$ in (24) and use the fact that the
pressure must vanish when the particles are infinitely heavy:
$$
P = -\int_0^{\infty} dm^2 \int {d^4 q\over (2\pi)^4} { \epsilon (q_0)
\over e^{\beta q^0}-1}\;{\rm Im}
{1\over q^2 - m^2 - \Pi (q, T, m)} \eqno (37c)
$$
where $\Pi$ is the self
energy defined in (24b) and $\epsilon (q_0)=\pm 1$
according to whether $q_0$ is positive or negative.  
As any divergence of $\Pi$ now
appears in the denominator, the summation has made it harmless.  One
has to worry about the denominator possibly vanishing when $q=0$ and
$m=0$, but this will be rendered harmless by the $q^3$ 
appearing in $d^4q = q^3 dq d\Omega$.
The quantitities in (37) are all unrenormalised; there is still some work
needed, involving the theory of composite operators, to cast the formula
into one that involves only finite renormalised quantitities\defref\otto
{
D Boedeker, P V Landshoff, O Nachtmann and A Rebhan,
hep-ph/9806514
}.

\bigskip
{\bf Linear response theory}

Suppose that the thermal equilibrium of a plasma is disturbed by the
switching on at $t=0$ of an external electrostatic potential
$A^0_{{\rm ext}} (x)$.  Then the system's Hamiltonian acquires an extra
term
$$
H' (t) = \theta (t)\int d^3 x\, J^0 (x) A^0_{\rm ext} (x) \eqno (38)
$$
where $J^0$ is the charge density.  This will cause $J^0$ to change.
As it is a Heisenberg-picture operator, its equation of motion is
$$
{\partial J^0 \over \partial t} = i \left [ H + H', J^0 \right ] \eqno (39a)
$$
where $H$ is the original Hamiltonian.  When we take the thermal
average of this equation, the contribution from $H$ will disappear
because originally there was thermal equilibrium.  So the integrated
change in $\langle J^0 (x) \rangle$ at very large time is
$$
\delta \langle J^0 (x) \rangle = \int d^4 x' \,G_{R} (x-x')
A^0_{{\rm ext}} (x') \eqno (39b)
$$
where 
$$
G_R (x-x') = \theta (t-t') \langle [ J^0 (x), J^0 (x') ] \rangle \eqno
(39c)
$$
Taking the Fourier transform,
$$
\delta \langle J^0 (k) \rangle = G_R (k) \,A^0_{\rm ext} (k) \eqno (39d)
$$
We may express the retarded Green's function $G_R$ in terms of 
elements of the matrix Green's function ${\bf G}^{\mu\nu}$
$$
G_R = {\half} (G_{11}^{00} - G_{22}^{00} + G_{21}^{00} - G_{12}^{00}) 
\eqno (40)
$$
Each of the terms here may be calculated from pertubation theory.
However, it is simpler to express $G_R$ in terms of the 
function $G^{00}(k)$ that appears in the diagonal matrix associated
with ${\bf G}^{00}$ (see (23)):
$$
G_R = n(k^0) e^{\half\beta |k^0 |} \{ 
(G^{00}+G^{00*})\sinh \half\beta |k^0 | + (G^{00}-G^{00*})\sinh \half\beta k^0 \} \eqno (41a)
$$
On the other hand
$$
G_{11} = n(k^0)e^{\half \beta |k^0 |} \{ 
 (G^{00}+G^{00*})\sinh\half\beta |k^0 |
+ (G^{00}-G^{00*})\cosh\half\beta k^0  \}
\eqno(41b)
$$

So it is sufficient to calculate $G_{11}^{00}$ and change its imaginary
part by multiplying it by $\tanh {1\over 2}\beta k^0$.

\bigskip
{\sevenrm This work was supported by the
EC Programme ``Training and Mobility of Researchers", Network
``Hadronic Physics with High Energy Electromagnetic Probes",
contract ERB FMRX-CT96-0008, and by PPARC}

\bigskip\goodbreak
\medskip\immediate\closeout\rfile\writestoppt
\baselineskip=10pt{{\bf References}}\bigskip{\frenchspacing%
\parindent=20pt\escapechar=` \input refs.tmp\bigskip}\nonfrenchspacing
\bye